# Electric Field Induced Macroscopic Cellular Phase of Nanoparticles

Abigail Rendos+,[1] Wenhan Cao+,[2] Margaret Chern,[1] Marco Lauricella,[3] Sauro Succi,[3] Jörg G. Werner,[1,2] Allison M. Dennis,[1,4] Keith A. Brown,[1,2,5]

[1] Division of Materials Science & Engineering, Boston University, Boston, MA, USA.

[2] Department of Mechanical Engineering, Boston University, Boston, MA, USA.

[3] Center for Life Nano Science, Italian Institute of Technology, Rome, Italy.

[4] Biomedical Engineering Department, Boston University, Boston, MA, USA.

[5] Physics Department, Boston University, Boston, MA, USA.

+ Contributed Equally
*Corresponding Author: Keith A. Brown, email address: brownka@bu.edu

A suspension of nanoparticles with very low volume fraction is found to assemble into a macroscopic cellular phase under the collective influence of AC and DC voltages. Systematic study of this phase transition shows that it was the result of electrophoretic assembly into a two-dimensional configuration followed by spinodal decomposition into particle-rich walls and particle-poor cells mediated principally by electrohydrodynamic flow. This mechanistic understanding reveals two characteristics needed for a cellular phase to form, namely 1) a system that is considered two dimensional and 2) short-range attractive, long-range repulsive interparticle interactions. In addition to determining the mechanism underpinning the formation of the cellular phase, this work presents a method to reversibly assemble microscale continuous structures out of nanoscale particles in a manner that may enable the creation of materials that impact diverse fields including energy storage and filtration.

## Introduction

Electric fields provide a flexible means to manipulate soft matter, however, they interact with materials in electrolytic solutions through numerous distinct phenomena making the outcome of field-directed assembly difficult to predict. When a suspension of polarizable particles experiences a spatially uniform electric field, their induced dipoles lead them to form isolated chains along field lines, causing solidification through the electrorheological effect.[1–3] In contrast, suspensions have also been experimentally observed to assemble into macroscopic porous structures with particle-rich walls and particle-poor voids,[4–8] even though this phase has not been completely recapitulated in simulation.[3,9,10] While these porous structures suggest a path to realizing continuous mesoporous solids with extremely low densities, the origin of this phase is not clear even though, and perhaps because, the cellular phase was observed in vastly different systems spanning orders of magnitude in particle size, particle volume fraction, and electric excitation. While electrohydrodynamic (EHD) flow was identified as the primary mechanism of formation in two instances[4,5] and electroosmotic (EO) flow in another,[8] the other two examples list the interactions between induced dipoles as the origin of the structure.[6,7] Overall, the lack of a cohesive and encompassing explanation for the formation of the cellular phase hinders the ability to design macroscopic porous structures.

Here, we find that a nanoparticle suspension exhibits a macroscopic cellular phase when an AC voltage $V_{AC}$ and DC voltage $V_{DC}$ are simultaneously applied, despite using quantum dot (QD) particles with a diameter and volume fraction both at least an order of magnitude smaller than all prior examples of the cellular phase. Indeed, systematic study revealed that the cellular phase only formed in the presence of both $V_{DC}$ and $V_{AC}$ at volume fraction-dependent critical voltages. The complex interactions required to produce a cellular phase in this system include (1) electrochemistry to generate a DC current, (2) electrophoresis to aggregate particles into a 2D arrangement on one electrode, and (3) an instability driven by the long-ranged repulsive and short-ranged attractive EHD flow that nucleates at regions on the electrode with high local field enhancement. Notably, EO and other purely attractive interactions are competitive with the cellular phase and instead drive the system towards a cluster-phase (i.e. pearl chaining). This mechanistic explanation was compared to all previous examples of the electrically-mediated cellular phase to identify a set of unifying factors that appear to always be present, namely that the system adopts an effective 2D arrangement and features an in-plane interaction that is short-range attractive and long-range repulsive. This understanding paves the way towards the concerted formation of hierarchical porous structures that may impact fields including energy storage and filtration.[11,12]

**Experimental Methods**

As a model system for assembly, poly(maleic anhydride-alt-1-octadecene) (PMAO)-coated CdSe/CdS quantum dots (QDs),[13,14] were suspended in dilute borate buffer (3.125 mM, 5.5 nm Debye screening length)15 at a volume fraction $\varphi$ = 6×10-5 which is equivalent to a 25 nM particle concentration. For a typical assembly experiment, indium tin oxide (ITO) slides (2277 - University Wafer, 703176 – Sigma Aldrich) were prepared by sonicating them in acetone and subsequently isopropanol for 5 min each before drying them under an N2 stream. Finally, the ITO slides were placed into the 3D printed frame pictured in Fig. S2(a). A laser-cut polyimide spacer (2271K72 – McMaster) with a height fixed by a 177 ± 1 µm was then placed onto one of the ITO slides and 4 µL of the suspension was pipetted onto the (ITO)-coated glass and covered with a second ITO-coated glass slide to form a fluid cell as shown in Fig. 1(a).

This complete cell was then transferred to an Olympus BX43 microscope with a GS3-U3-120S6M-C Grasshopper camera. A filter cube with an emission wavelength at 642 nm, 75 nm BW (67-036 – Edmund Optics Inc.), a short-pass excitation filter with a cutoff at 500 nm (84-706 – Edmund Optics Inc.), and a dichroic with cutoff at 550 nm (DMLP550R – ThorLabs Inc.) were used to visualize the photoluminescent QDs. Alligator-clip leads were attached to a corner of each ITO-coated slide as shown in Fig. S2(b) in order to apply an AC voltage with amplitude $V_{AC}$ and DC offset voltage $V_{DC}$ across the fluid cell using a Keysight 33521B waveform generator. Fluorescence micrographs were taken using 5× magnification with a 500 ms exposure time and 14 dB gain to have sufficient contrast. In most experiments, such as those conducted in Figs. 1 and 5 along with supplemental Figs. S3, S4, and S7, $V_{AC}$ = 2 V with frequency $f$ = 500 kHz and $V_{DC}$ = 2.2 V.

**Results and Discussion**

Simultaneously applying $V_{AC}$ and $V_{DC}$ across the QD suspension resulted in a cellular phase at strikingly low $\varphi$, particle size, and field intensities. Specifically, setting $V_{DC}$ = 2.2 V and $V_{AC}$ = 2 V amplitude at 500 kHz, the particles assembled into a cellular phase over the course of a few minutes as shown in Fig. 1(b). To determine whether this process was reversible, the field was subsequently switched off, which led the suspension to gradually homogenize through diffusion, as seen in Fig. 1(c). Given that this phase has not been previously observed for particles with hydrodynamic diameters < 100 nm, we considered whether this could be specific to these QDs. Thus, we repeated this experiment with commercially available fluorophore-doped polystyrene nanoparticles and again observed the cellular phase (Fig. S3), showing that this phase is not restricted to these QDs.

To explore the mechanism of the cellular phase and whether it originated from forces between induced dipoles, we compute the non-dimensional parameter $\Lambda = \frac{\alpha V^2}{8 k_B T H^2}$, which reflects the importance of induced dipole interactions between particles relative to thermal energy given particle polarizability $\alpha$, applied voltage $V$, electrode separation $H$, Boltzmann's constant $k_B$, and temperature $T$. Here, we estimate $\Lambda \sim 0.008$, indicating that induced dipole mediated assembly should not occur and that other interactions must drive the formation of the cellular phase. Another potential mechanism is suggested by the resemblance of the cellular phase to Benárd cells where gravity-driven natural convection from density gradients produces similar cells.[16–18] Thus, we repeated the experimental conditions shown in Fig. 1 with the cell rotated 90° such that gravity pointed along the electrodes and the same cellular structure formed (Fig. S4), indicating that natural convection is not responsible for the cellular phase.

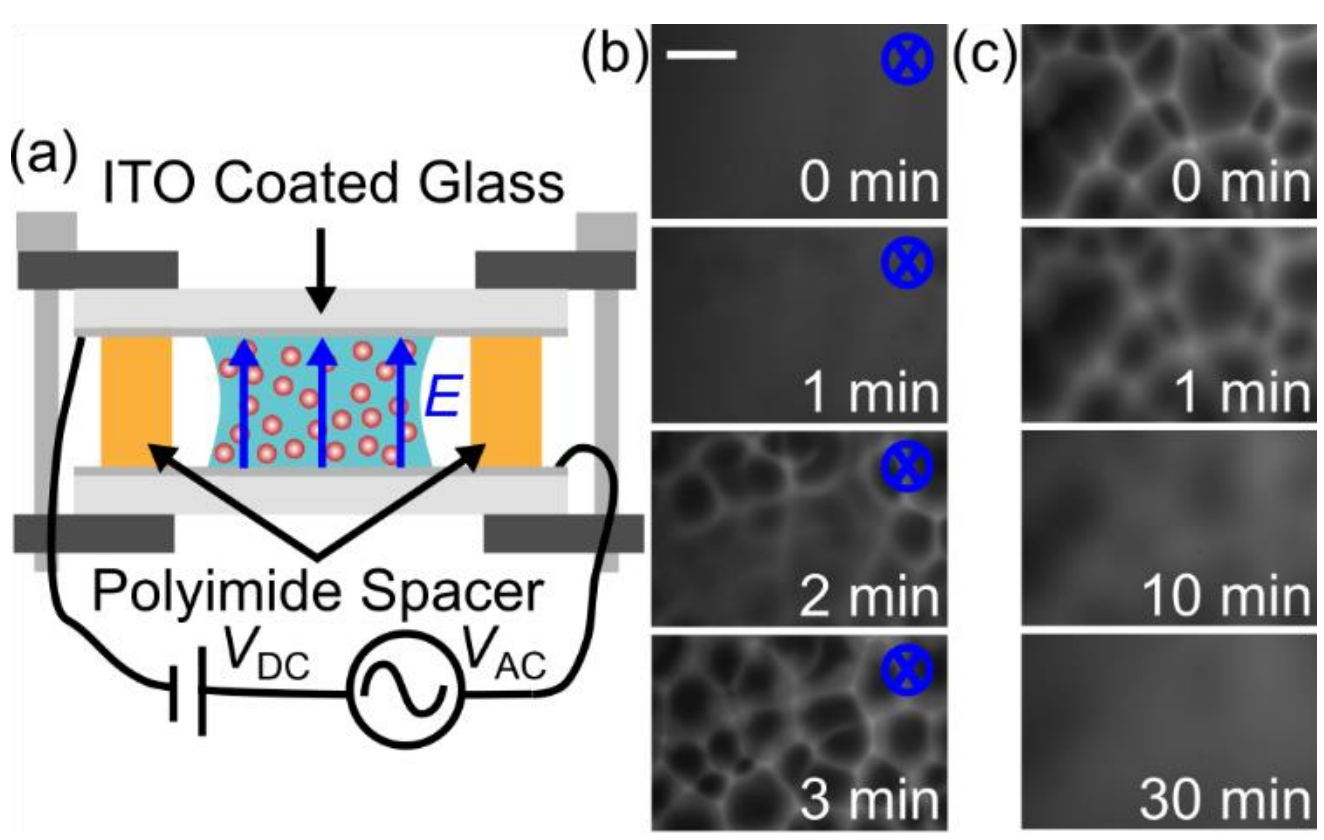

**Fig. 1** (a) Fluid cell consisting of two ITO-coated glass slides separated by a polyimide spacer and containing a suspension of quantum dots (QDs). DC and AC voltages, $V_{DC}$ and $V_{AC}$ respectively, were applied across the fluid cell. (b) Fluorescent micrographs showing the formation of the cellular phase when $V_{DC}$ = 2.2 V and $V_{AC}$ = 2 V were applied to a suspension with volume fraction $\varphi = 6\times10^{-5}$. The scale bar depicts 500 µm. (c) Fluorescent micrographs showing that after $V_{DC}$ and $V_{AC}$ were turned off, the particles diffused back into a homogenous distribution.

In order to explore the mechanism of the cellular phase, we examined the contributions of $V_{DC}$ and $V_{AC}$. Specifically, we performed a series of experiments holding $V_{AC}$ fixed while $V_{DC}$ was incrementally increased from 0 to 2.6 V in steps of 0.2 V. A fluorescence micrograph was taken at each increment after equilibrating for 4 min. The observed gradual transition from a uniform suspension to an ordered cellular phase was apparent and shown in Fig. 2(a). To analyze these experiments, the fluorescent pattern in each image was manually classified as having (1) a uniform background (no phase) (2) an interconnected series of lines (cellular phase), (3) a series of isolated bright spots (cluster phase) or (4) indeterminate (transition phase). These classifications are shown in Fig. 2(b), which clearly shows that a minimum $V_{DC}$ of ~1.8 V was required to form the cellular phase. The full series of images are shown in Fig. S5.

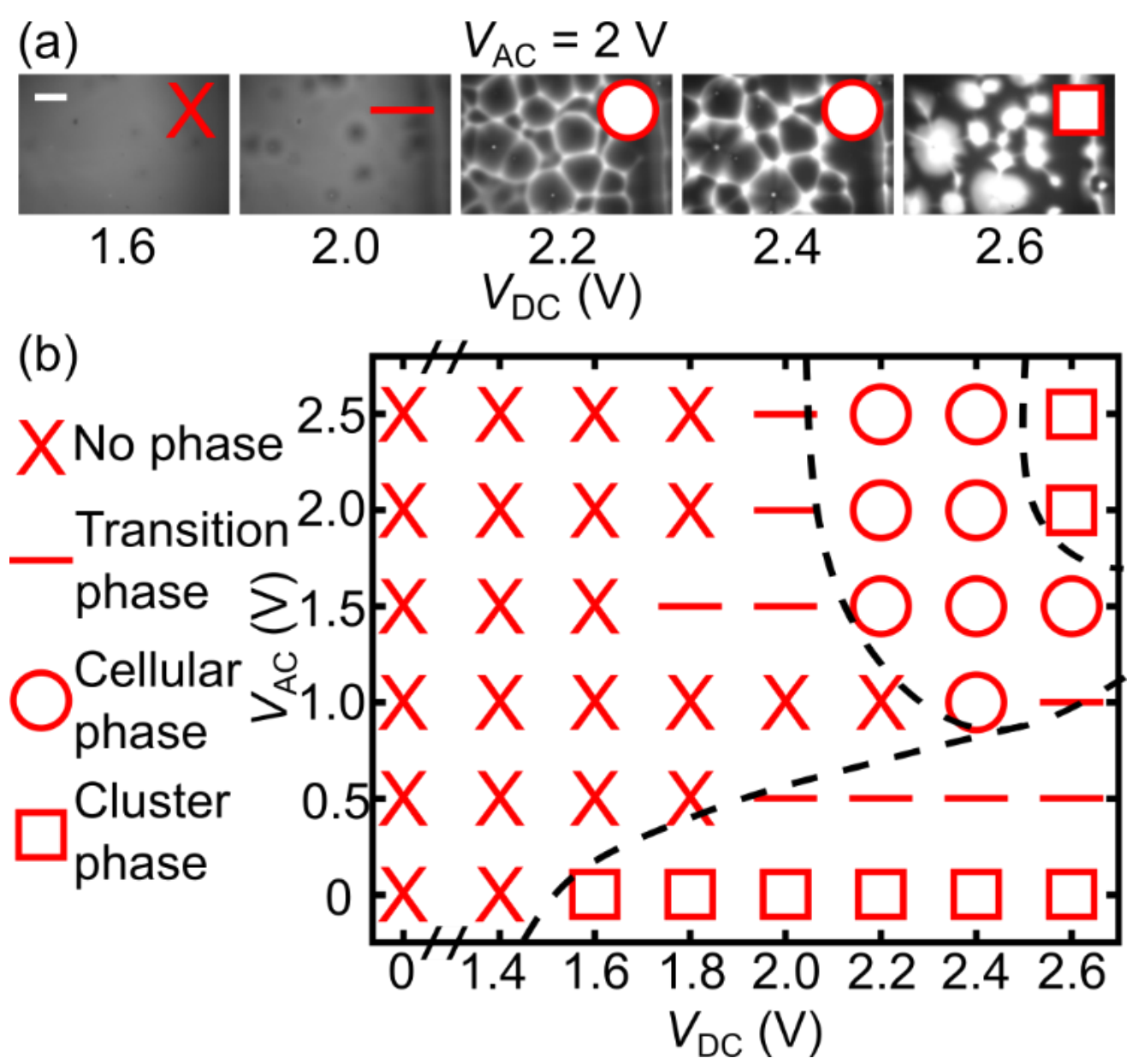

**Fig. 2** (a) Phase transitions as $V_{DC}$ was increased with $V_{AC}$ = 2.0 V for $\varphi$ = 6×10$^{-5}$. The scale bar depicts 500 μm. (b) Phase diagram depicting transition from no phase to either cellular or cluster phases. Ambiguous images were classified as transition. Full data is shown in Fig. S5.

While $V_{DC}$ was required to form the cellular phase, the origin of the ~1.8 V threshold voltage, or the dominant physical effect of $V_{DC}$, was not clear. The relevance of the DC field is especially noteworthy when one considers that electrolyte ions will accumulate on oppositely charged electrodes, giving rise to an electrode polarization that screens the DC field, thus preventing a strong DC component in the bulk.[19,20] One process that could maintain a steady-state DC field is a constant flow of ions across the chamber mediated by their electrochemical generation/annihilation at the anode and cathode. Electrochemical reactions at the electrodes could also explain the $V_{DC}$ threshold as highly non-linear currents are common in electrochemistry due to reaction-specific standard reduction/oxidation potentials and mass transport effects.[21] To determine whether electrochemical currents were present, two-electrode cyclic voltammetry measurements were conducted on the cell, confirming that electrochemical reactions were present and resulted in an appreciable current when $V_{DC} > 1.5$ V (Fig. S6). The observed reactions were likely due to a combination of water electrolysis and ITO degradation.[22,23] While the current turn-on behavior was commensurate with the onset of the cellular phase, we sought to establish a more definitive link between electrochemical reactions on the electrodes and the cellular phase. Thus, we coated the surface of the ITO electrodes with an insulating layer of poly(methyl methacrylate) (PMMA) to prevent electrochemical reactions from occurring and repeated the assembly experiment. After this treatment, the cellular phase did not form at any voltage (Fig. S7), verifying that electrochemistry was required for the cellular phase to form.

Due to the electrochemical reactions at the electrodes, a steady-state DC field persisted in the chamber and led to electrophoresis of the QDs and a subsequent increase in their local concentration on one electrode. Since these QDs had a slightly negative zeta potential,[25] they will accumulate on the positively charged electrode and form a thin particle-dense film, calculated to be about a monolayer in thickness. To understand the fate of this initially uniform film, it is useful to consider that the magnitude of $V_{AC}$ determines whether it will adopt a cellular or cluster phase, as shown in Fig. 2(b). Particles on a substrate will interact through two types of electrically-induced fluid flows: EO flow and EHD flow.[21,25,26] While EO flow is a DC phenomenon, EHD flows arise from both $V_{AC}$ and $V_{DC}$.[20,24] Interestingly, these effects are expected to produce contrasting flow fields in which EO draws particles together in the plane while EHD flow, despite being short-range attractive, will repel particles at long ranges. While a Cahn-Hilliard analysis of these interactions revealed that both interactions can drive a spinodal decomposition, and will do so in a concentration-dependent manner, the instigator of the cellular phase was not clear from this analysis alone.

To determine what interaction drove the spinodal decomposition from a film to the cellular phase, we performed a series of experiments at various $\varphi$ in which $V_{DC}$ was held constant while $V_{AC}$ was gradually increased. First, we prepared a sample with $\varphi$ = 3×10-5, $V_{DC}$ = 1.9 V, and increased $V_{AC}$ from 0.5 to 5.5 V in steps of 0.5 V. To quantify the critical AC voltage $V_{AC}^{*}$ at which the cellular phase forms, the images were analyzed to count the number N of cells in each image (Fig. S8). Fitting $N$ vs. $V_{AC}$ to a sigmoid (Eq. S16) allowed us to quantify $V_{AC}^{*}$. A typical experiment is shown in Fig. 3(a). Eight conditions were tested in triplicate (at four values of $\varphi$ and both $V_{DC}$ = 1.9 V and $V_{DC}$ = 2.2 V) over the range of $V_{AC}$, enabling a Cahn-Hilliard analysis of the cellular phase formation. In this framework, spinodal decomposition is predicted to occur when the interparticle interaction (i.e. EO or EHD) leads perturbations to grow faster than they dissipate through diffusion. Due to these competing effects, a general relationship is expected wherein the strength of EHD and EO are assigned unknown, but concentration and field independent, pre-factors $\beta_{EHD}$ and $\beta_{EO}$. EHD and EO scale with electric field quadratically and linearly, respectively.[21] Thus, the data in Fig. 3(b) was fit to,

$$\varphi^{-1} = \beta_{EHD}(V_{AC}^{*} + bV_{DC})^{2} + \beta_{EO}V_{DC} \qquad (1)$$

where b reflects that while AC and DC voltages can both give rise to EHD, they may have different intensities.[21] Using nonlinear least squares fitting, we found $\beta_{EHD}$ = 0.004 ± 0.001 V-2, $b$ = 9.7

± 0.8, and $\beta_{EO}$ = -0.9 ± 0.2 V$^{-1}$. Critically, since both $\beta_{EHD}$ and $b$ were positive, this means that EHD promoted the formation of the cellular phase. In contrast, $\beta_{EO}$ being negative means that EO flow inhibited the cellular phase formation. Interestingly, these results implied that VDC played two competing roles by contributing to both EO and EHD. These results demonstrated that EHD flow was critical to the formation of the cellular phase.

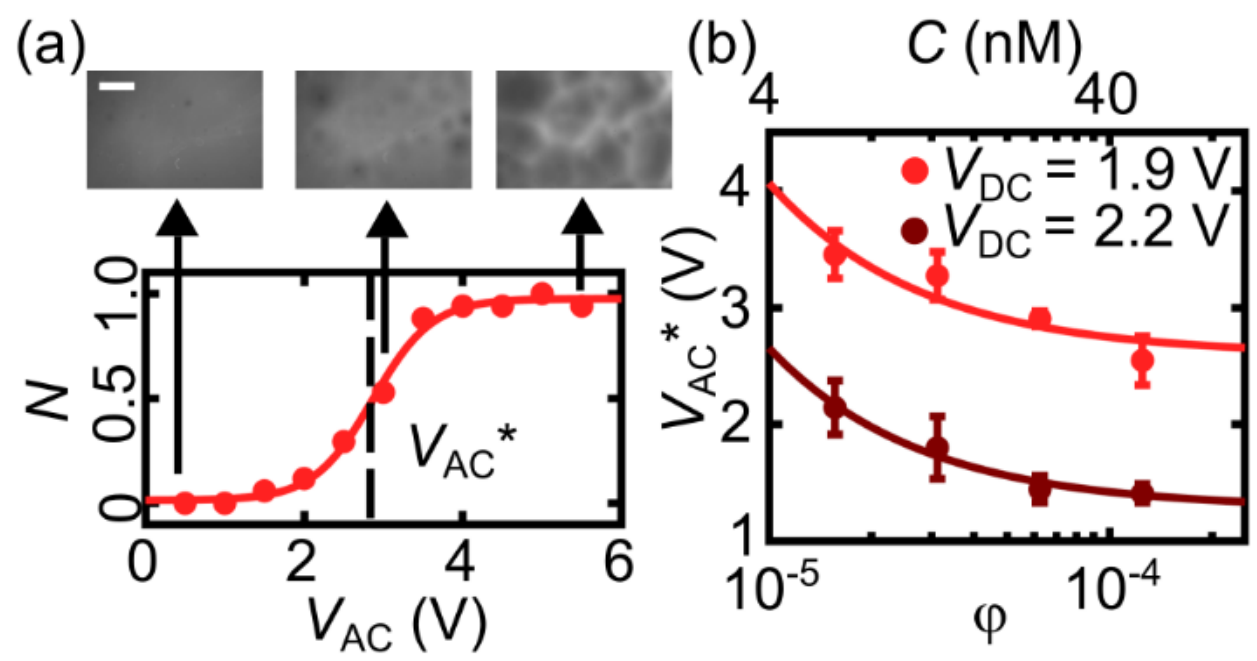

**Fig. 3** (a) Using image processing, the normalized number $N$ of cells for $\varphi$ = 3×10$^{-5}$ and $V_{DC}$ = 1.9 V vs. $V_{AC}$ are plotted and fit to a sigmoid to identify the transition voltage $V_{AC}^*$. Representative images displayed with a 500 μm scale bar. (b) $V_{AC}^*$ vs. $\varphi$ and concentration $C$ for $V_{DC}$ = 1.9 and 2.2 V with the lines depicting a single fit to Eq. 1.

Given that EHD was identified as an instigator, and EO an inhibitor, of the cellular phase, other features of assembly can be understood by considering where the cells form. Specifically, by performing the assembly experiment with conditions that lead to the cellular phase ($V_{DC}$ = 2.2 V, $V_{AC}$ = 2.0 V, $\varphi$ = 6×10-5), leaving the field off for 40 min to allow the particles to homogenize, and then repeating the same experiment, we found that the structure of the cellular phase was repeatable with voids occurring at the same locations as shown in Fig. S9(a). To explore this further, an experiment was performed where a solution was exposed to conditions that led to a cellular phase ($V_{DC}$ = 2.2 V, $V_{AC}$ = 3.0 V, $\varphi$ = 6×10-5), the system was then allowed to homogenize with the field off, and then exposed to conditions that led to a cluster phase ($V_{DC}$ = 2.4 V, $V_{AC}$ = 0 V). Importantly, Fig. S9(b) shows that many of the cluster phases were co-localized with the centers of the voids of the cellular phase. Together, these results suggest that features of the underlying substrate, likely asperities that enhance local electric field,[28] break the symmetry of the system and nucleate the phase transition. Furthermore, the fact that the same location can lead to voids through repulsive EHD flows or clusters through attractive EO flows further suggests that the mode of spinodal decomposition is fundamentally different between EHD- and EO-mediated phases.

These experiments and analysis coalesced into a proposed mechanism for the cellular phase formation involving electrophoresis, EO flow, and EHD flow. Once $V_{DC}$ was applied to the particle suspension, electrochemistry at the electrodes led to a DC current that electrophoretically pulled the particles to one side of the chamber as shown in Fig. 4(a). Once assembled into a film, EO led to an attractive flow that promoted particle aggregation as shown in Fig. 4(b). However, $V_{AC}$ and $V_{DC}$ also produced an EHD flow as depicted in Fig. 4(c) that was short-range attractive but repulsive at long ranges. Both flow profiles in Fig. 4(d) were replotted from Ristenpart *et al.*[21] Depending on which flow dominated, spinodal decomposition in Fig. 4(e) either began through the nucleation of an excess or decrease of particles at the high field regions, which subsequently led to the cluster phase or cellular phase, respectively. A similar dichotomy of spinodal decompositions has been observed in simulations of colloids with competing interparticle interactions.[29,30] Interestingly, at $V_{DC} > 2.4$ V, a transition from a cellular phase to a cluster phase was observed, but this is qualitatively different than the low $V_{AC}$ cluster phase as it occurs at the nodes of the cells. Thus, we attribute this to the vertices of the cells becoming tall enough to span the chamber, at the expense of the structure becoming thinner, at which point particles are recirculated into the voids.

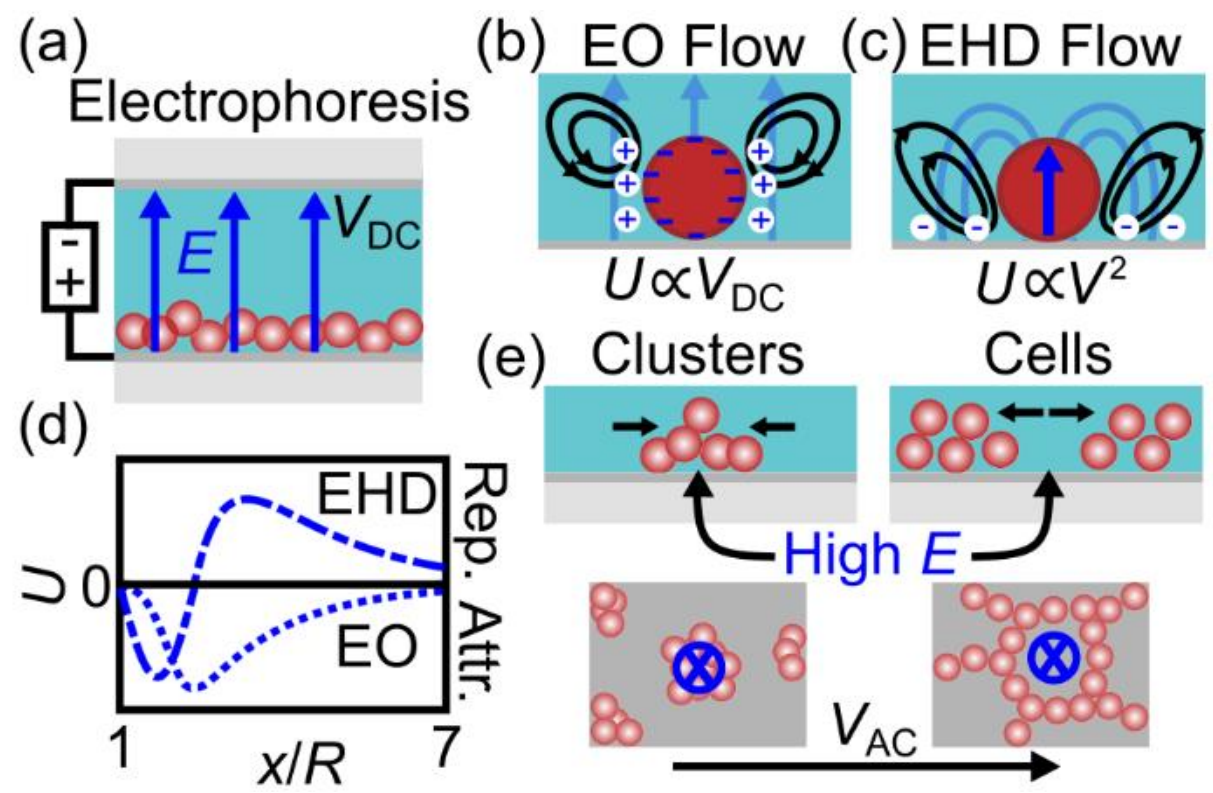

**Fig. 4** Proposed mechanism of cellular phase formation. (a) With application of $V_{DC}$, electrophoresis moved the QDs to one electrode. (b) Once assembled into a film, electroosmotic (EO) flow produced circulating flows that promote aggregation. (c) Simultaneously, electrohydrodynamic (EHD) flow promoted aggregation at short distances, but is otherwise repulsive. (d) Flow velocity $U$ replotted from Ristenpart *et al*.[22] vs. distance $x$ from a particle of radius $R$ for both EO (dotted line) and EHD (dashed line). (e) For small $V_{AC}$, EO flow dominated and particles aggregated at high field regions causing cluster formation. For larger $V_{AC}$, EHD dominated and produced voids at the same high field regions, nucleating the cellular phase.

With a greater understanding of the mechanism of the cellular formation, we hypothesized that the cell arrangement could be controlled. Having shown that a polymer coating prevented the formation of the cellular phase, we reasoned that photoresist could serve as a patternable coating to localize assembly. To test this, we patterned a star on the ITO slide connected to the positive lead and performed an assembly experiment. After applying $V_{AC}$ = 2 V and $V_{DC}$ = 2.2 V for 4 min, the cellular structure formed with cells only present centered at the points of the star (Fig. 5). Interestingly, this simple method allowed for the location and orientation of five cells to be controlled, suggesting further opportunities for crafting complex macro-porous arrangements with very low particle densities.

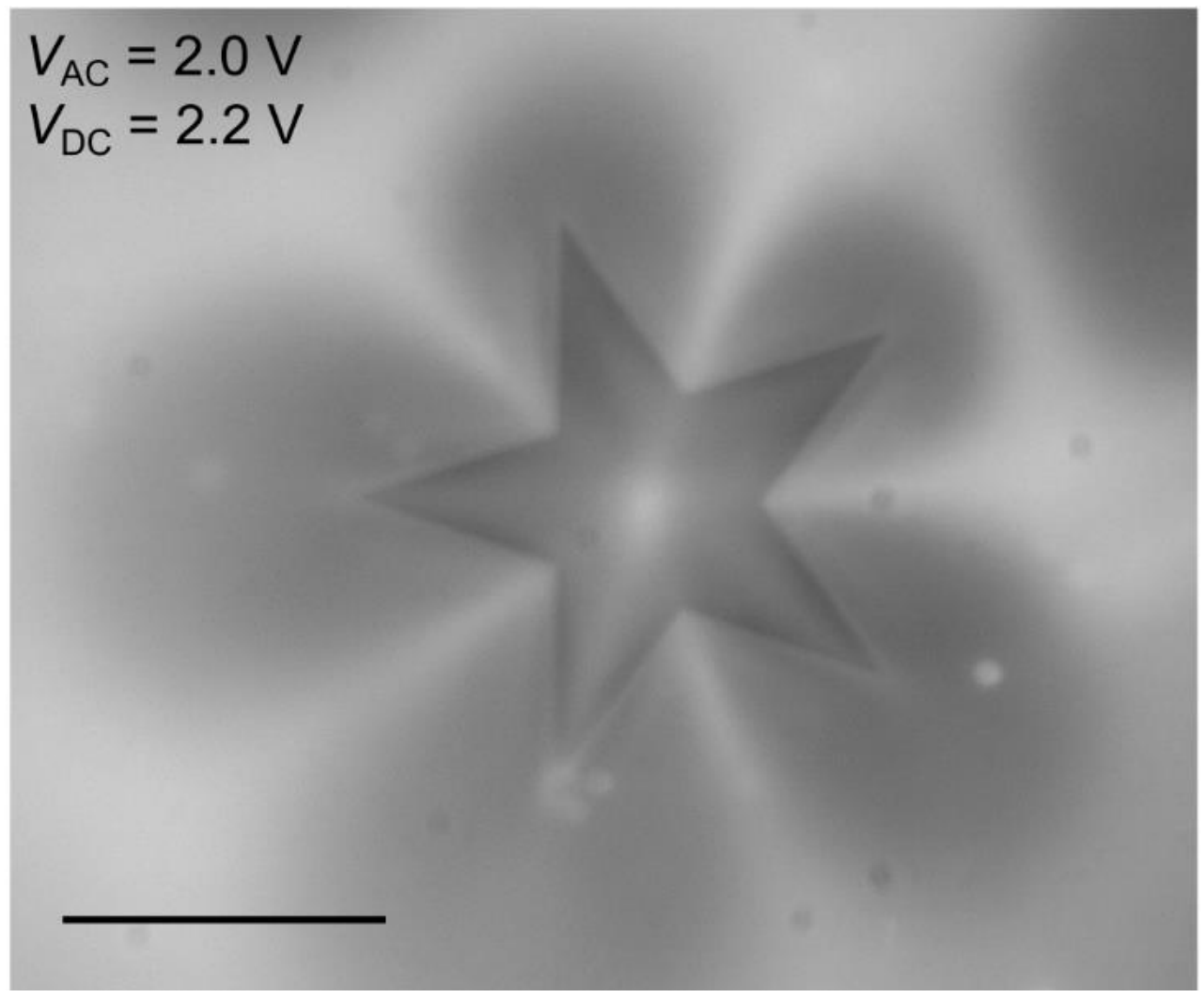

**Fig. 5** Fluorescence micrograph of cells formed at the points of the star pattern where the star is the exposed ITO and the rest is covered with photoresist. Here, $V_{AC}$ = 2.0 V, $V_{DC}$ = 2.2 V, and $\varphi = 6\times10^{-5}$. The scale bar depicts 500 μm.

Considering the mechanism of formation identified in the present work and the characteristics of prior work that resulted in the cellular phase, two commonalities emerge that unify all observations of the electrically mediated cellular phase. The first commonality is that each system can be effectively reduced to 2D by way of gravity or $V_{DC}$ pulling particles to one side of the chamber,[4,5,8] or through the formation of chains spanning the entire chamber.[6,7] The second commonality is that all feature interactions that are short-range attractive and long-range repulsive in the plane of the electrode, either through dipolar interactions of particle chains or through EHD flow. Indeed, a qualitatively similar cellular phase has been observed in suspensions of magnetic particles under the influence of triaxial magnetic fields that were effectively 2D systems in which the interaction was short-range attractive and long-range repulsive,[31–33] suggesting that these features can contribute towards a more general and complete understanding of the cellular phase.

## Conclusion

We observed an ultra-low density macroscopic cellular phase through the electrically mediated assembly of nanoparticles that were an order of magnitude smaller than previous examples. Additional control experiments helped tease out the factors that contribute to cellular phase formation such as EO and EHD flows and the importance of electrochemistry at the electrode surface. This interaction between electrochemistry, electrophoresis, EO, and EHD results in a unique porous structure made of nanoparticles where the characteristic length of the pores is 10,000 times larger than the size of the particles. Importantly, by comparing to prior work, we identified two characteristics that appear to be required to form the cellular phase: particles that are confined in some way to 2D and an interparticle interaction that is short-range attractive, long-range repulsive. This level of understanding is essential to bridging the gap between observing and utilizing the unique structures produced by this assembly process.

## Acknowledgements

Acknowledgment is made to the donors of the American Chemical Society Petroleum Research Fund for partial support of this research under award 57452-DNI9. A.R. acknowledges support from the Boston University Nanotechnology Innovation Center. We acknowledge support from the Boston University Photonics Center, the College of Engineering Dean's Catalyst Award, and the Gordon and Betty Moore Foundation.

# Supplemental Information
# Electric Field Induced Macroscopic Cellular Phase of Nanoparticles

Abigail Rendos+,[1] Wenhan Cao+,[2] Margaret Chern,[1] Marco Lauricella,[3] Sauro Succi,[3] Jörg G. Werner,[1,2] Allison M. Dennis,[1,4] Keith A. Brown,[1,2,5]

[1] Division of Materials Science & Engineering, Boston University, Boston, MA, USA.

[2] Department of Mechanical Engineering, Boston University, Boston, MA, USA.

[3] Center for Life Nano Science, Italian Institute of Technology, Genova, Italy.

[4] Biomedical Engineering Department, Boston University, Boston, MA, USA.

[5] Physics Department, Boston University, Boston, MA, USA.

+Contributed Equally

*Corresponding Author: Keith A. Brown, email address: brownka@bu.edu

## I. Cahn-Hilliard Analysis

To analyze the spinodal decomposition of the particles from a uniform distribution to a cellular phase, we perform a Cahn-Hilliard analysis. As an initial state for this analysis, we posit that the DC electric field $E$ will lead the particles to assemble onto the electrode that is positively charged. To justify this, we estimate the electrophoretic speed $v_p$ of the QDs which is given by,[1]

$$v_p = \frac{\varepsilon_m \zeta_p E}{\eta} \text{ for } \kappa R \gg 1, \tag{S1}$$

with particle radius $R$, inverse Debye length $\kappa$, particle zeta potential $\zeta_p$, medium permittivity $\varepsilon_m$, and medium viscosity $\eta$. For our system of QDs suspended in 3.125 mM borate buffer, $\kappa$ = 0.18 nm$^{-1}$ and R = 8.5 nm, so $\kappa R > 1$. Based on Eq. S1, a particle with $\zeta_p \cong$ -30 mV[2] at room temperature in water will move ~0.2 mm/s when 2 V is applied across 200 µm. Under these conditions, the particle would traverse the fluid cell in ~1 s, so all the particles will concentrate on the electrode immediately upon application of $E$. If the QDs have a packing fraction of 0.74 and are dispersed at a bulk volume fraction $\varphi$ of 6×10$^{-5}$, they are expected to form a film ~12 nm thick on the surface of the electrode which is approximately a monolayer of particles

Once assembled into a two-dimensional film, the movement of particles can be described by the convection-diffusion equation,[3]

$$\frac{\partial n(\boldsymbol{r},t)}{\partial t} + \nabla \cdot [n(\boldsymbol{r},t)\boldsymbol{U}(\boldsymbol{r},t)] = D\nabla^2 n(\boldsymbol{r},t), \tag{S2}$$

where the first term describes the change in particle areal concentration $n$ with time $t$ at a location $\boldsymbol{r}$ on the surface of the electrode, the second term describes the convection of particles due to flow field $\boldsymbol{U}$, and the final term represents the diffusion of the particles with diffusion coefficient $D$, which is assumed to be constant. The total number of particles is not changing and as a result no source or sink term is included.

After assembly onto one electrode due to electrophoresis, the initial areal concentration $n_0$ on the positively charged electrode is given by,

$$n_0 = \frac{3\varphi_0 H}{4\pi R^3}, \tag{S3}$$

where $\varphi_0$ is the bulk volume fraction and $H$ is the chamber height. In the Cahn-Hilliard analysis, a plane wave perturbation $n'$ is added and the growth or decay of this term will determine the stability of the film. This leads to an expression,

$$n(\mathbf{r},t) = n_0 + a(t)\mathrm{e}^{ikx}, \tag{S4}$$

where $a(t)$ is the amplitude of the wave perturbation, $k$ is the non-dimensional wave vector normalized by $R$, and $x$ is a direction along the electrode. We assume that initially the perturbation $n'$ is small compared to $n_0$. Due to this perturbation, the flow field can be separated into $\boldsymbol{U} = \boldsymbol{U}_0 + \boldsymbol{U}'$ where $\boldsymbol{U}_0$ is due to $n_0$ and $\boldsymbol{U}'$ is due $n'$. However, we assume the initially uniform distribution of particles indicates that $\boldsymbol{U}_0 = 0$. Thus, Eq. S2 can be linearized to show,

$$\frac{\partial n'}{\partial t} + n_0 \nabla \cdot \boldsymbol{U}' = D\nabla^2 n'. \tag{S5}$$

By introducing Eq. S4 into Eq. S5 and simplifying, the expression becomes,

$$e^{ikx}a'(t) + \; n_0\nabla\cdot\boldsymbol{U'} = Da(t)\nabla^2 e^{ikx}. \tag{S6}$$

Drawing from the analysis performed by Hardt *et al*,[4] we define $\boldsymbol{U'}$ generally using the integral,

$$U' = \frac{-\mathrm{a(t)}e^{ikx}i}{4\pi\eta}\int_{-\pi}^{\pi}\int_{0}^{\infty}\sin(krcos\theta)\cos(\theta)\cdot v(r)\,drd\theta, \tag{S7}$$

where $v(r)$ is the flow velocity as a function of the magnitude of the location $r$ in the x-y plane and $\theta$ describes the direction of $r$. Importantly, there were two flow types present in our system, electroosmotic (EO) and electrohydrodynamic (EHD) flow. The EHD velocity $v_{EHD}(r)$ and EO velocity $v_{EO}(r)$ can be defined as,

$$v_{EHD}(r) = V^2\gamma_{EHD}f_{EHD}(r) \tag{S8}$$

and

$$v_{EO}(r) = V_{DC}\gamma_{EO}f_{EO}(r), \tag{S9}$$

where the functions $f(r)$ describe the flow at a point $r$ away from a single particle based on the flow profiles in Fig. 4(d) which were replotted from theory described by Ristenpart *et al*.[5] and the constants $\gamma$ describe the strength of the flow field with subscripts denoting EHD and EO flow. It is known that EHD flow is proportional to the voltage squared, whereas EO flow scales directly with the applied voltage[5] as described in Eqs. S8 and S9. For each flow, the integral in Eq. S7 was solved numerically as a function of $k$. Thus, Eq. S7 can be expressed as,

$$U'(k) = \frac{-a(t)e^{ikx}i}{4\pi\eta}(V^2\gamma_{EHD}W_{EHD}(k) + V_{DC}\gamma_{EO}W_{EO}(k)), \tag{S10}$$

where $W_{EHD}(k)$ and $W_{EO}(k)$ are the result of the double integral described by Eq. S7 for $f_{EHD}$ and $f_{EO}$, respectively. The plot of $W(k)/k$ in Fig. S1 for EHD shows that at low $k$ (or high wavelength $\lambda$) repulsion between particles is expected.

To apply this understanding to our data, Eq. S10 was introduced into Eq. S6 and simplified,

$$a'(t) = a(t)(-Dk^2 + \; \frac{kn_0}{4\pi\eta}(V^2\gamma_{EHD}W_{EHD}(k) + V_{DC}\gamma_{EO}W_{EO}(k))), \tag{S11}$$

which describes the evolution of the amplitude over time. When the term multiplied by $a(t)$ is positive, the perturbation grows, leading to spinodal decomposition. Thus, Eq. S11, can be used to compute a critical voltage $V^*$ at which an instability will occur which is found to be,

$$V^* = \sqrt{\frac{4\pi\eta Dk}{n_0\gamma_{EHD}W_{EHD}(k)} - \frac{\gamma_{EO}W_{EO}(k)V_{DC}}{\gamma_{EHD}W_{EHD}(k)}}. \tag{S12}$$

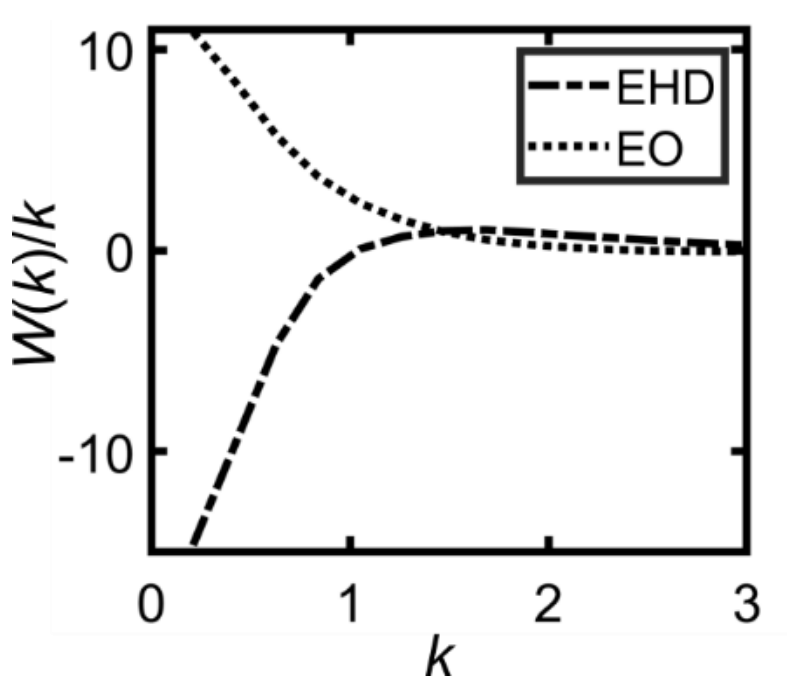

**Fig. S1** $W(k)/k$ plotted versus wavenumber $k$ for both EHD and EO flows.

Using the Stokes-Einstein equation for $D$,[6]

$$D = \frac{k_B T}{6\pi\eta R}, \tag{S13}$$

where $k_B$ is Boltzmann's constant and $T$ is temperature, Eq. S12 can be simplified to,

$$V^* = \sqrt{\frac{2k_B Tk}{3Rn_0\gamma_{EHD}W_{EHD}(k)} - \frac{\gamma_{EO}W_{EO}(k)V_{DC}}{\gamma_{EHD}W_{EHD}(k)}}. \tag{S14}$$

Importantly, Eq. S14 shows that $V^*$ is inversely related to $n_0$, which implies that as the volume fraction increases, the necessary voltage to observe the cellular phase will decrease. Additionally, $V^*$ must be expanded into its AC and DC components,

$$V^* = V^*_{AC} + bV_{DC}, \tag{S15}$$

where the dimensionless constant $b$ allows $V_{AC}$ and $V_{DC}$ to contribute to EHD with intensities reflecting the different complex conductivities at DC and high frequencies.7 In our experiments, the critical AC voltage $V^*_{AC}$ at which the cellular phase was observed was determined by fitting the experimental data to a sigmoid as seen in Fig. 3(a),

$$\frac{N}{N_{max}} = a_1 + \frac{a_2}{1+e^{a_3(V_{AC}-V^*_{AC})}}, \tag{S16}$$

where $N$ is the number of cells, $N_{max}$ is the maximum number of cells observed in that experiment, and $a_1$, $a_2$, and $a_3$ are additional fitting parameters. Incorporating Eq. S15 and S3 into Eq. S14 yields,

$$V_{AC}^{*} = \sqrt{\frac{8\pi k k_B T R^2}{9H\varphi_0 \gamma_{EHD} W_{EHD}(k)} - \frac{\gamma_{EO} W_{EO}(k) V_{DC}}{\gamma_{EHD} W_{EHD}(k)}} - bV_{DC}, \tag{S17}$$

which was simplified to the functional form used to fit the data in Fig. 3(b),

$$V_{AC}^{*} = \sqrt{\frac{1}{\beta_{EHD}\varphi_0} - \frac{\beta_{EO} V_{DC}}{\beta_{EHD}}} - bV_{DC}. \tag{S18}$$

where $b$, $\beta_{EHD}$, and $\beta_{EO}$ are fitting parameters. Equation S18 captures the behavior in our experimental determination of $V_{AC}^{*}$ shown in Fig. 3(b) and confirms that as $V_{AC}^{*}$ increases, $\varphi_0$ and $V_{DC}$ decrease.

## II. Supplementary Tables and Figures

TABLE S1. Summary of experimental conditions and materials used for previous studies of the cellular phase and this current work.

| Paper | Particle type | Particle diameter | Particle volume fraction | Medium | Field strength |
|---|---|---|---|---|---|
| Trau – 1995[7] | Barium titanate | 100 nm | 0.025 vol% | Castor oil | DC: 0.05 V/µm |
| Sapozhnikov – 2003[8] | Bronze spheres | 120 (& 40) µm | 3 vol% | Ethanol toluene mixture | DC: 0.66 V/µm |
| Kumar – 2005[9] | poly-alpha olefin spheres | 45 to 87 µm | 0.5 – 10 vol% | Corn oil | AC: 0.1 to 3 kHz, 1.6 – 5 V/µm |
| Agarwal – 2009[10] | Silica | 800 nm | 0.07 – 4 vol% | Water + dimethyl sulfoxide mixture | AC: 1 MHz, 1 V/µm |
| Hardt – 2020[4] | DNA | ~350 nm[11] (hydrodynamic diameter) | ~0.02 vol% | Dextran and polyethylene glycol | DC: 10-30 V, length across which field is applied is not specified |
| This work | QDs | 17 nm (hydrodynamic diameter)[12] | 0.0015 – 0.012 vol% | 3.125 mM borate buffer | Typical conditions: DC: 0.01 V/µm AC: 500 kHz, 0.01 V/µm |

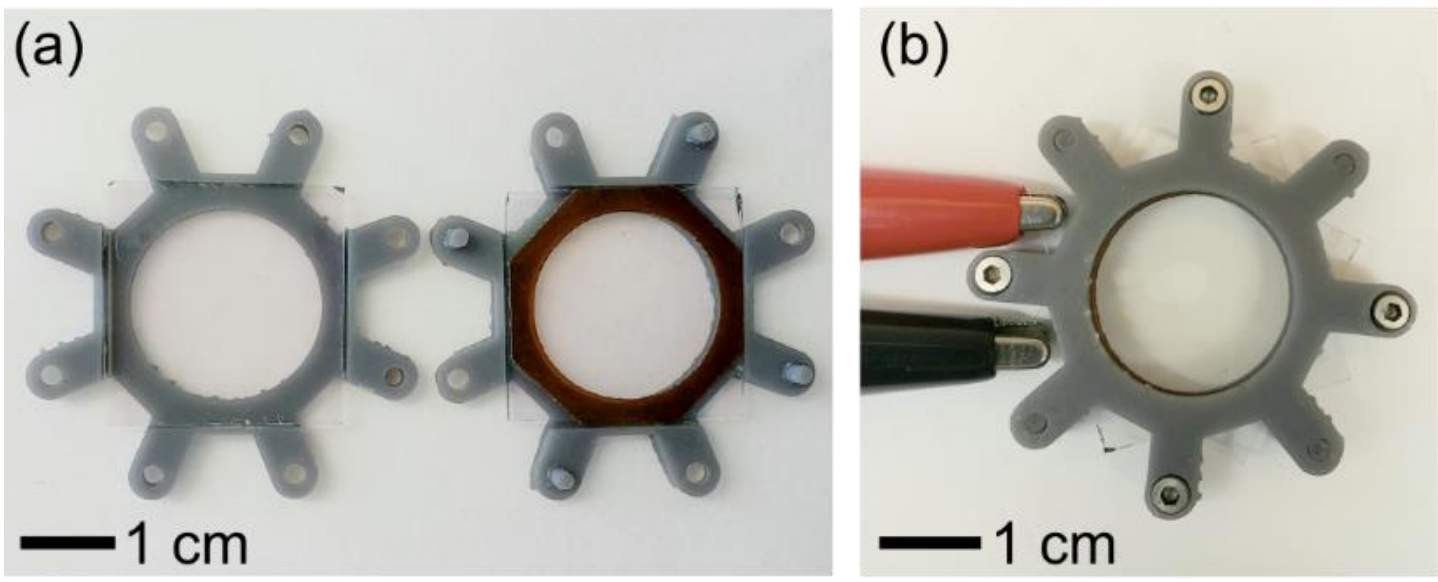

**Fig S2.** (a) Top and bottom of 3D printed fluid cell frame designed to hold one-inch square indium tin oxide (ITO)-coated glass slides. ITO-glass slides are shown with the polyimide spacer on one slide. Arms were used to align the cell and screw cell together. (b) Fully assembled fluid cell with glass slides, droplet, spacer, screws, and leads.

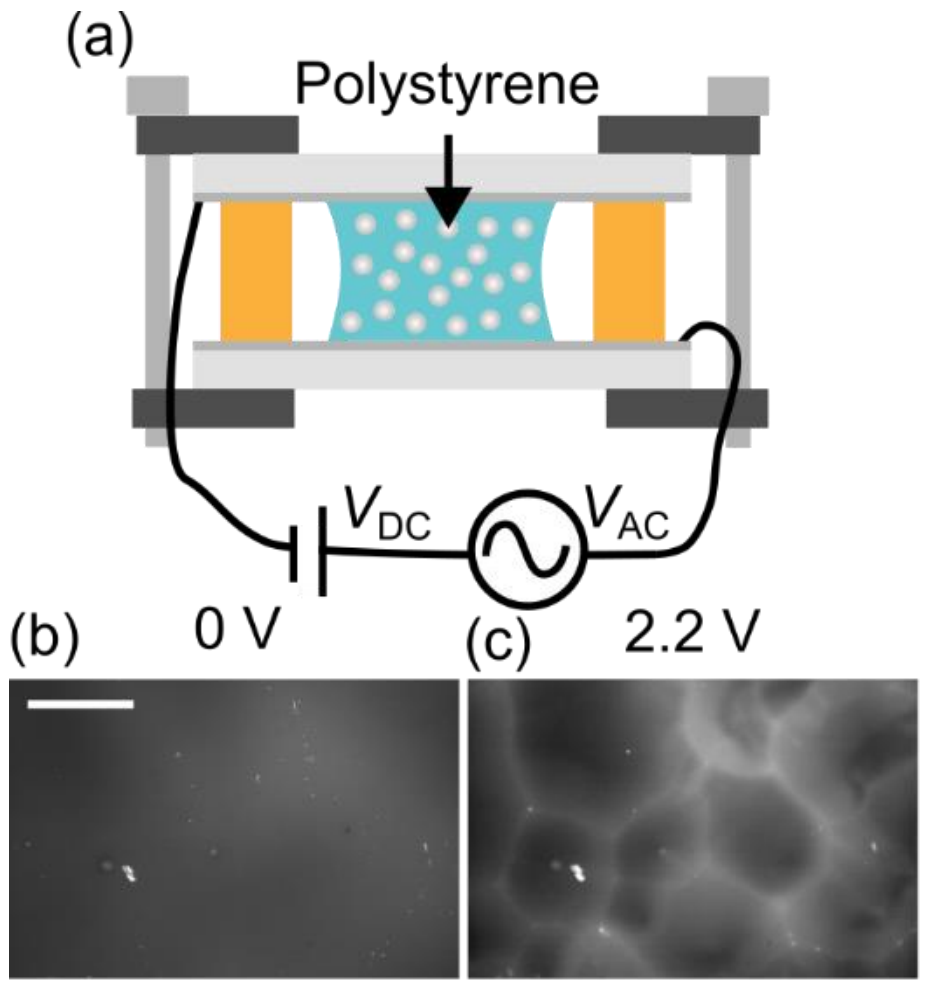

**Fig S3.** (a) Experimental set-up for fluid cell with polystyrene particles with a diameter of 26 nm at a volume fraction $\varphi = 9\times10^{-5}$. (b) Fluorescence micrograph of cell when field was off. Scale bar depicts 500 μm. (c) Image of cell after applying AC voltage amplitude $V_{AC} = 2$ V with frequency $f = 500$ kHz and turning on DC voltage $V_{DC} = 2.2$ V for 4 min.

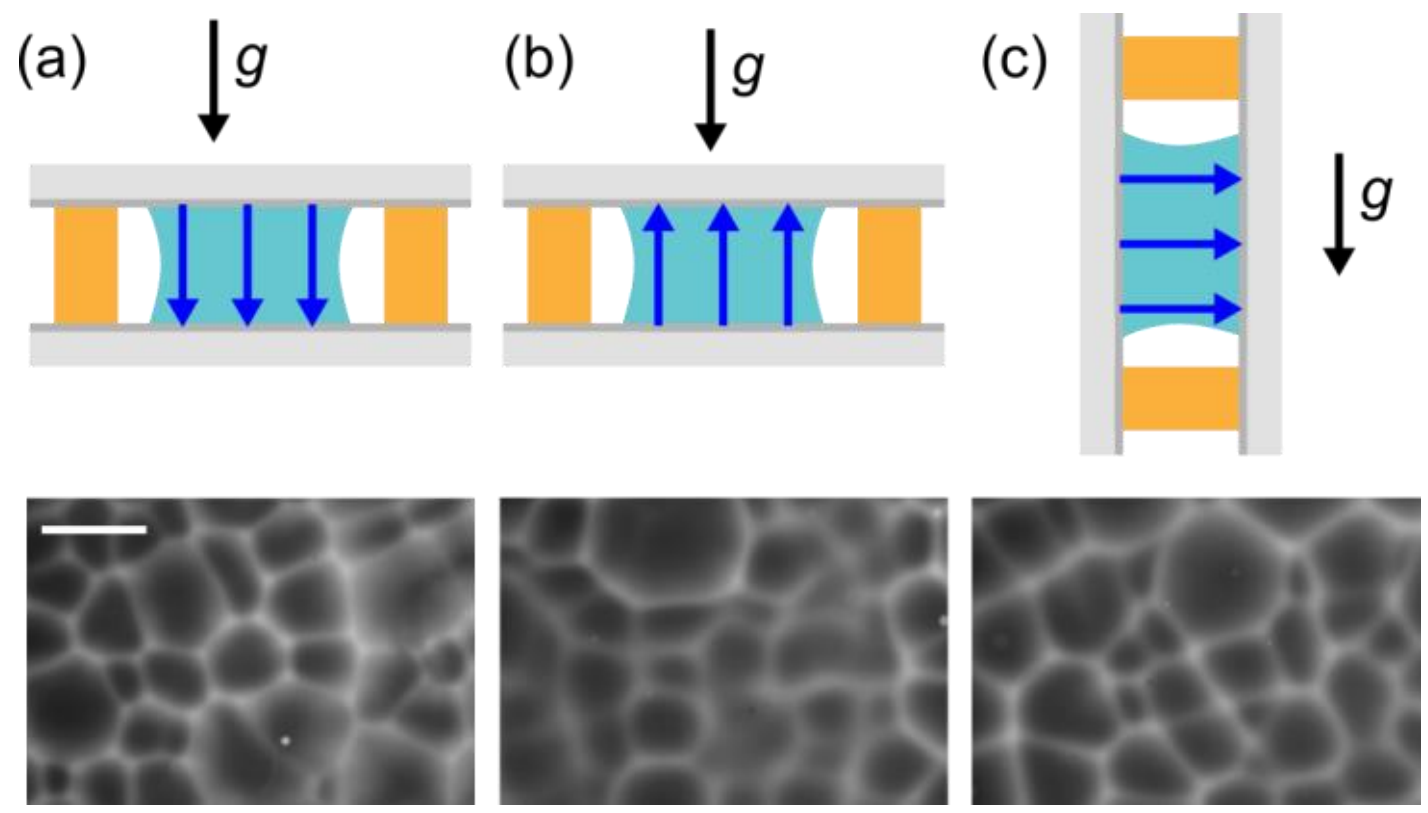

**Fig S4.** Buoyancy experiments performed with the cell in three orientations relative to gravity $g$: (a) $g$ and applied electric field $E$ in alignment (b) $g$ and $E$ in opposite directions and (c) $g$ and $E$ perpendicular to one another. Each shows a fluorescence micrograph of the cellular phase at $V_{AC}$ = 2 V and $V_{DC}$ = 2.2 V with QDs at $\varphi = 6\times10^{-5}$ where the QDs accumulate on the positive electrode. The scale bar depicts 500 μm.

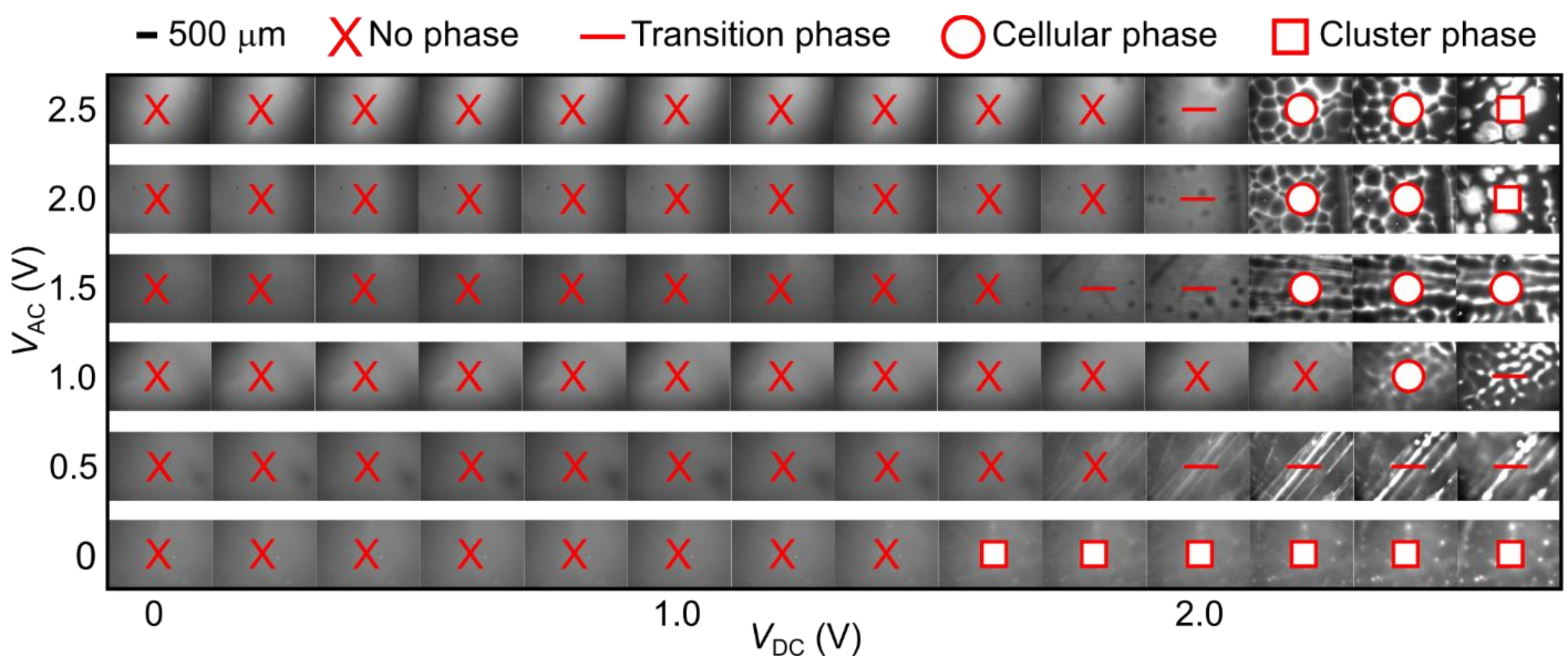

**Fig S5.** Full set of fluorescence micrographs of assembly experiments of QDs at $\varphi = 6\times10^{-5}$ in which $V_{AC}$ was initially set to a value from 0 V to 2.5 V and then $V_{DC}$ was increased from 0 V to 2.6 V in steps of 0.2 V. After each time $V_{DC}$ was increased, the system was allowed to stabilize for 4 min before images were taken.

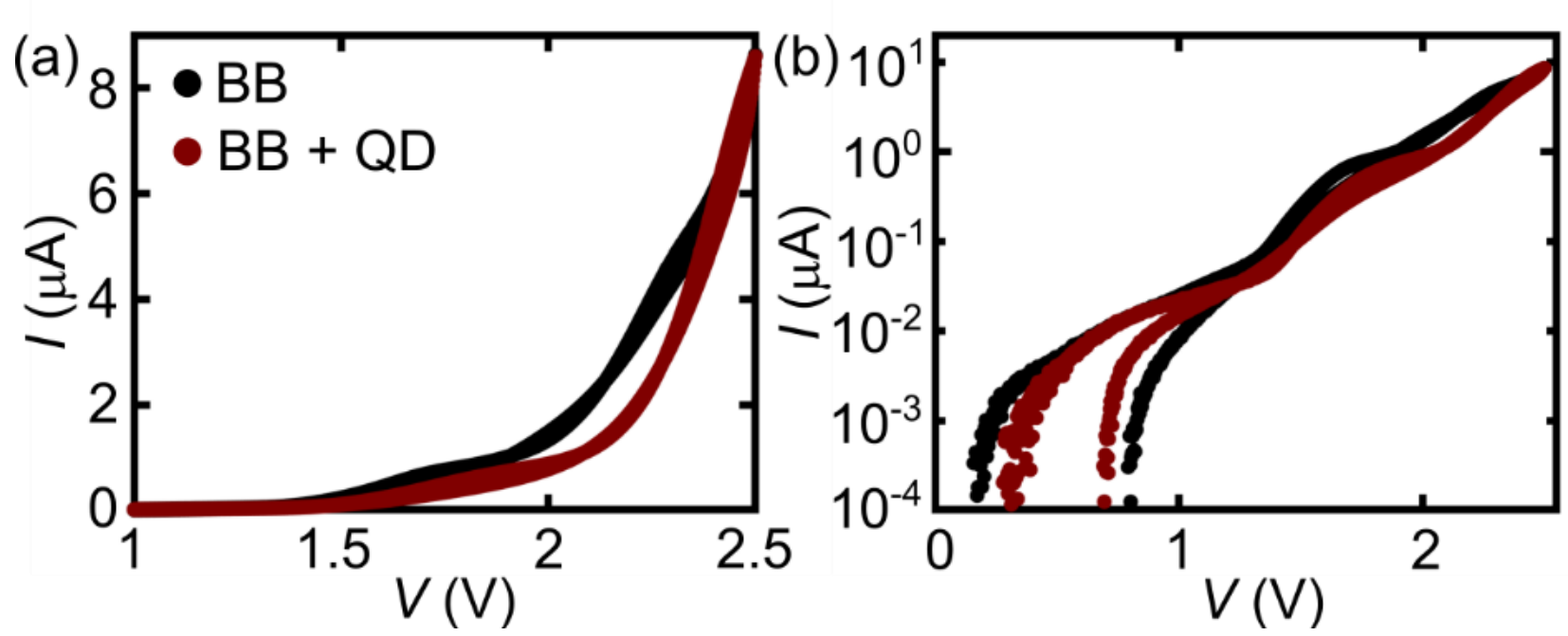

**Fig S6.** Current $I$ versus voltage $V$ plots from two-electrode cyclic voltammetry electrochemical measurements conducted at a sweep rate of 20 mV/s using the Gamry Reference 600+ Potentiostat/Galvanostat/ZRA on the fluid cell with 3.125 mM borate buffer (BB) solution and QDs in 3.125 mM borate buffer (QD + BB) solution with $\varphi = 6\times10^{-5}$ in (a) linear and (b) log-y scale.

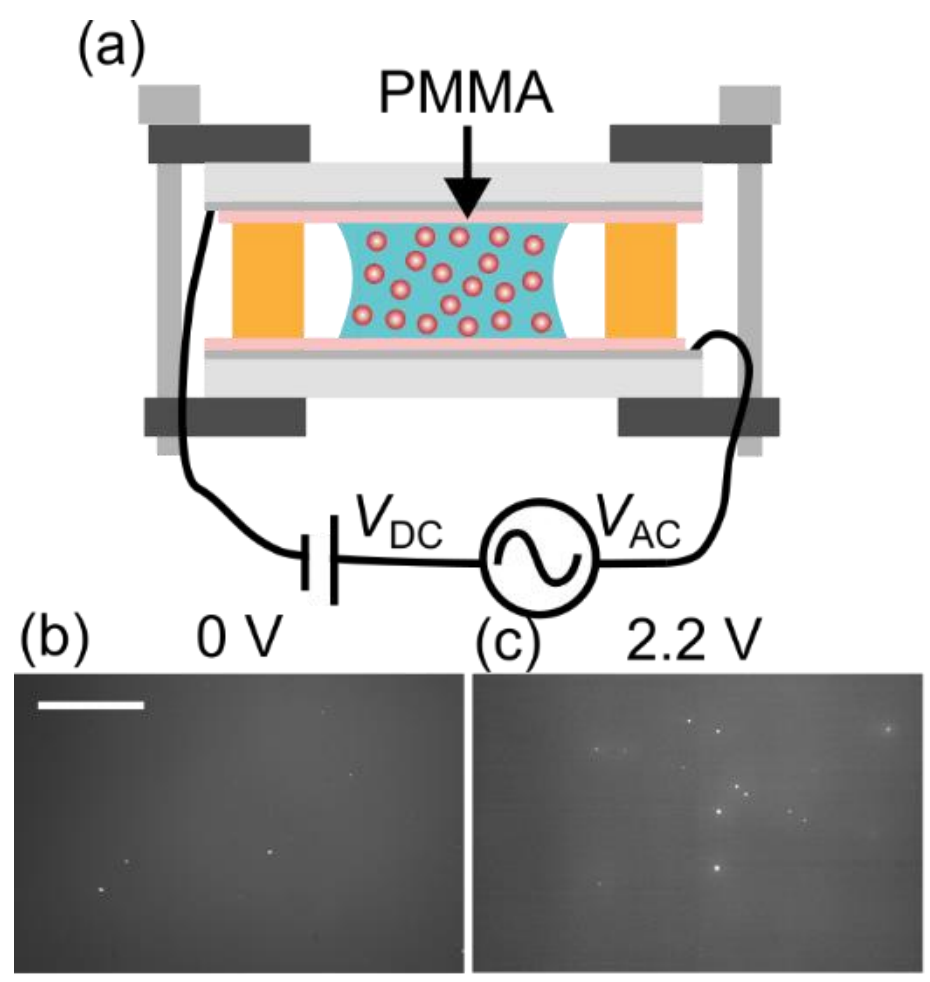

**Fig S7.** (a) Experimental set-up for fluid cell with polymethylmethacrylate (PMMA) coated ITO glass slides made by spin-coating PMMA (molecular weight = 950 kg/mol, diluted to 6% solids weight) onto both ITO glass slides and then baking them on a hot plate for 2 min at 100℃. (b) Images of top view when field was off with $\varphi = 6\times10^{-5}$. The scale bar depicts 500 μm. (c) Image after applying $V_{AC} = 2$ V and turning on $V_{DC}$ for 4 min.

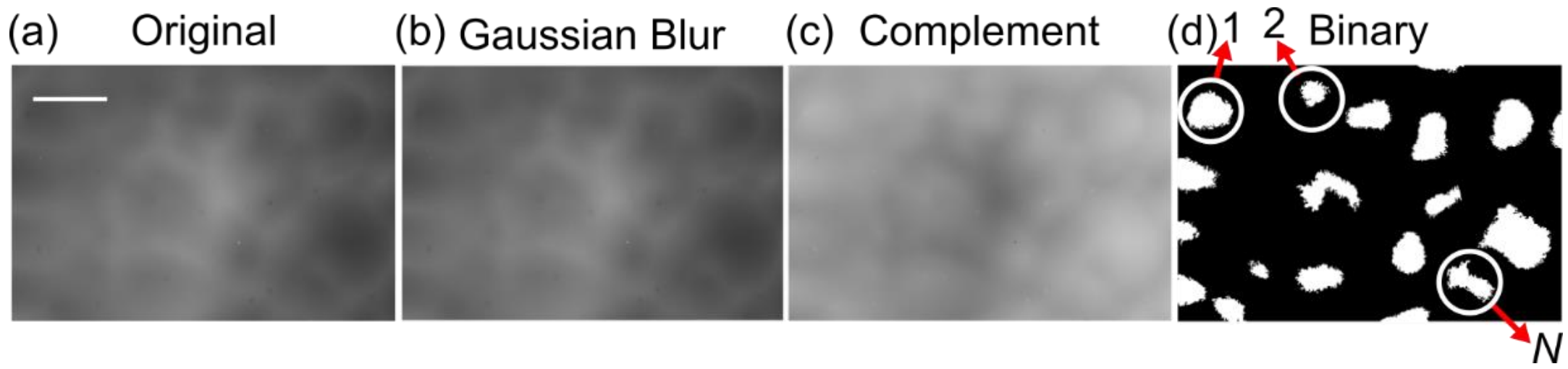

**Fig S8.** (a) Fluorescence microscopy image taken at 5× magnification of QDs with $\varphi$ = 3×10$^{-5}$, $V_{DC}$ = 1.9 V and $V_{AC}$ = 4 V. The scale bar depicts 500 μm. (b) Image after Gaussian blur was applied. (c) Complement of image, white spots indicate voids in the original image. (d) A binary map of the image enabling identification of the number of cells $N$.

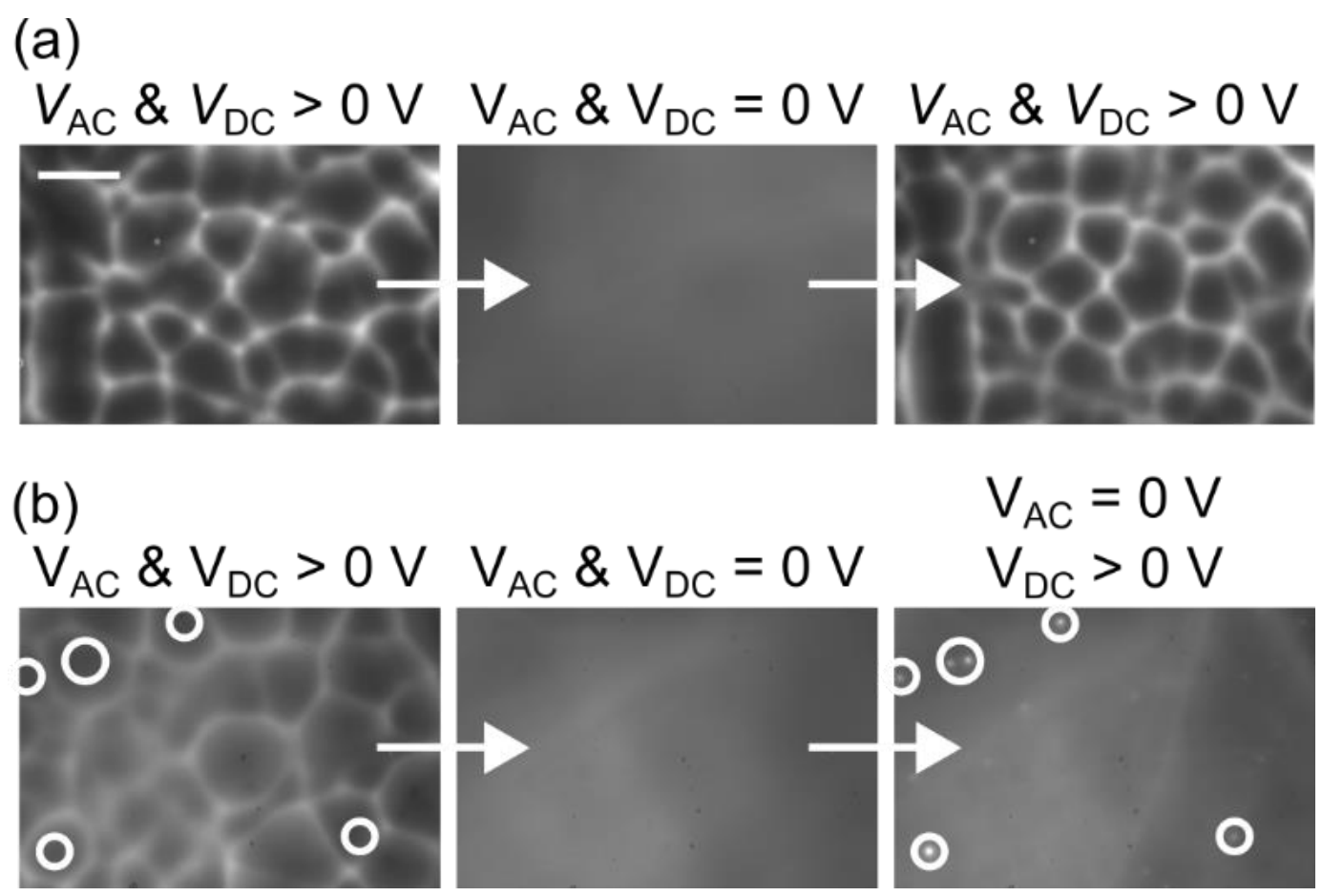

**Fig S9.** (a) From left to right, micrograph of QDs with $\varphi$ = 6×10$^{-5}$ when $V_{DC}$ = 2.2 V and $V_{AC}$ = 2 V after equilibrating for 4 min, then the field was turned off for 40 min to allow the particles to redistribute, and finally the same voltage was applied for 4 min. Scale bar depicts 500 μm and applies to both (a) and (b). (b) From left to right, micrograph of QDs at $\varphi$ = 6×10$^{-5}$ after $V_{DC}$ = 2.2 V and $V_{AC}$ = 3 V had been applied for 4 min, micrograph taken after 40 min after the field had been turned off, and then a micrograph taken after $V_{DC}$ = 2.4 V was subsequently applied for 4 min. The location of bright spots where QD concentration is high are indicated by the white circles, these same spots correspond to the location of voids in the left-most micrograph.